\documentclass[aps,prl,superscriptaddress,twocolumn,showpacs,amsmath,amssymb,floats]{revtex4}

\usepackage{txfonts}
\usepackage{amssymb}
\usepackage{graphicx}

\begin{document}

\title{Isostructural spin-density-wave and superconducting gap anisotropies in iron-arsenide superconductors}

\author{T. T. Han}
\affiliation{International Center for Quantum Materials, School of Physics, Peking University, Beijing 100871, China}

\author{L. Chen}
\affiliation{International Center for Quantum Materials, School of Physics, Peking University, Beijing 100871, China}

\author{C. Cai}
\affiliation{International Center for Quantum Materials, School of Physics, Peking University, Beijing 100871, China}

\author{Y. D. Wang}
\affiliation{International Center for Quantum Materials, School of Physics, Peking University, Beijing 100871, China}

\author{Z. G. Wang}
\affiliation{International Center for Quantum Materials, School of Physics, Peking University, Beijing 100871, China}

\author{Z. M. Xin}
\affiliation{International Center for Quantum Materials, School of Physics, Peking University, Beijing 100871, China}

\author{Y. Zhang}\email{yzhang85@pku.edu.cn}
\affiliation{International Center for Quantum Materials, School of Physics, Peking University, Beijing 100871, China}
\affiliation{Collaborative Innovation Center of Quantum Matter, Beijing 100871, China}

\date{\today}

\begin{abstract}
When passing through a phase transition, electronic system saves energy by opening energy gaps at the Fermi level. Delineating the energy gap anisotropy provides insights into the origin of the interactions that drive the phase transition. Here, we report the angle-resolved photoemission spectroscopy (ARPES) study on the detailed gap anisotropies in both the tetragonal magnetic and superconducting phases in Sr$_{1-x}$Na$_x$Fe$_2$As$_2$. First, we found that the spin-density-wave (SDW) gap is strongly anisotropic in the tetragonal magnetic phase. The gap magnitude correlates with the orbital character of Fermi surface closely. Second, we found that the SDW gap anisotropy is isostructural to the superconducting gap anisotropy regarding to the angular dependence, gap minima locations, and relative gap magnitudes. Our results indicate that the superconducting pairing interaction and magnetic interaction share the same origin. The intra-orbital scattering plays an important role in constructing these interactions resulting in the orbital-selective magnetism and superconductivity in iron-based superconductors.

\end{abstract}


\maketitle

High-$T_c$ superconductivity always intertwines with the symmetry breaking phases, whose origins are closely related to the pairing interaction of superconductivity. In the parent compounds of iron-based superconductors \cite{gap1,gap2,gap3,gap4}, a nematic phase first arises when the system breaks rotational symmetry through a structural transition from tetragonal to orthorhombic. A collinear antiferromagnetic phase (AFM/O) transition follows at a lower temperature, which further breaks the translational symmetry. By suppressing both the nematic and AFM/O phases through carrier doping or pressure, high-$T_c$ superconductivity emerges. Understanding the connections between superconductivity and symmetry breaking phases is thus crucial. It has been proposed that the superconducting pairing can be mediated by either the nematic or spin fluctuations \cite{gap5, gap6, gap7, gap8}.

Electronic system opens energy gaps in the symmetry breaking phases. The gap structure, especially the gap anisotropy in the momentum space usually reflects the detailed form of the interactions that drive the phase transition. For iron-based superconductors, the superconducting gap anisotropy has been intensively studied. The superconducting gap is isotropic in most materials \cite{gap9, gap10, gap11}, but anisotropic in FeSe, BaFe$_2$(As$_{1-x}$P$_x$)$_2$, and etc. \cite{gap12, gap13, gap14, gap15}. For the AFM/O phase, spin-density-wave (SDW) gap opens on the Fermi surface. Delineating the SDW gap anisotropy is crucial for understanding the microscopic origin of the magnetic interaction in iron-based superconductors. However, in contrast to the superconducting gap structure, the detailed delineation of the SDW gap structure is still lacking so far. One complexity comes from the two-fold rotational symmetry of the AFM/O phase. The band structure is complicated by the nematic band splitting and sample twinning effect \cite{gap16, gap17, gap18}.

\begin{figure}[t]
\includegraphics[width=8.7cm]{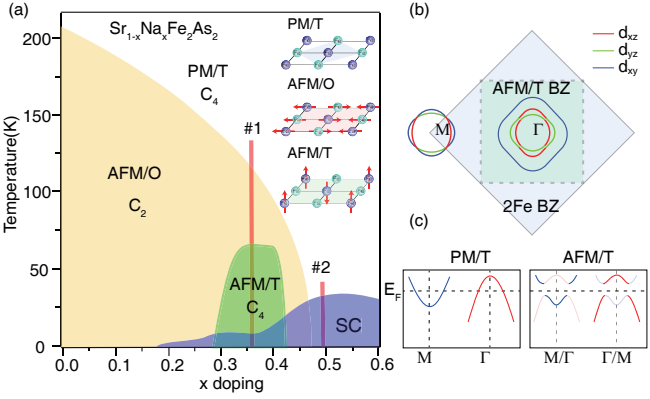}
\caption{(color online) Sr$_{1-x}$Na$_x$Fe$_2$As$_2$ phase diagram and schematic of the magnetic structure and band folding in the magnetic ordered phase. (a) The phase diagram adapted from Ref.\cite{gap20}. The red vertical lines show the doping levels of our samples. Schematic drawing of the lattice and magnetic structures in the paramagnetic (PM/T), collinear antiferromagnetic (AFM/O), and tetragonal magnetic (AFM/T) phases. The red arrows represent the directions of magnetic moments. (b) The general Fermi surface of iron-based superconductors. The orbital characters of Fermi surface are illustrated using different colors. (c) Schematic of the band folding between $\Gamma$ and M in the AFM/T phase. }\label{f1}
\end{figure}

Recently, a tetragonal magnetic phase (AFM/T) has been discovered in hole-doped iron-based superconductors \cite{gap19,gap20,gap21,gap22,gap23}. The AFM/T phase emerges in a small doping and temperature regime in the phase diagram \cite{gap20} [Fig.~\ref{f1}(a)]. Neutron scattering and M\"ossbauer spectroscopy show that the magnetic structure of the AFM/T phase is a double-Q SDW \cite{gap22, gap23}. It can be viewed as a superposition of ($\pi$, 0) and (0, $\pi$) collinear SDW with the magnetic moments rotating from in-plane to out-of-plane direction [Fig.~\ref{f1}(a)]. While the translational symmetry breaking persists, the rotational symmetry recovers from two-fold to four-fold in the AFM/T phase. Therefore, the AFM/T phase provides us with a good opportunity to study the SDW gap anisotropy in iron-based superconductors. Here, we studied the detailed gap structures in both the AFM/T and superconducting phases in Sr$_{1-x}$Na$_x$Fe$_2$As$_2$ using angle-resolved photoemission spectroscopy (ARPES). We found that the SDW and superconducting gap anisotropies are isostructural with similar angular dependencies, gap minima locations, and relative gap magnitudes. This establishes a close connection between the magnetism and superconductivity in iron-based superconductors, suggesting that the superconducting pairing is mediated by spin fluctuation. The pocket-dependence and orbital-selectivity of the gap anisotropy further highlight the important role of intra-orbital scattering in driving both the magnetism and superconductivity in iron-based superconductors.

\begin{figure*}[t]
\includegraphics[width=15cm]{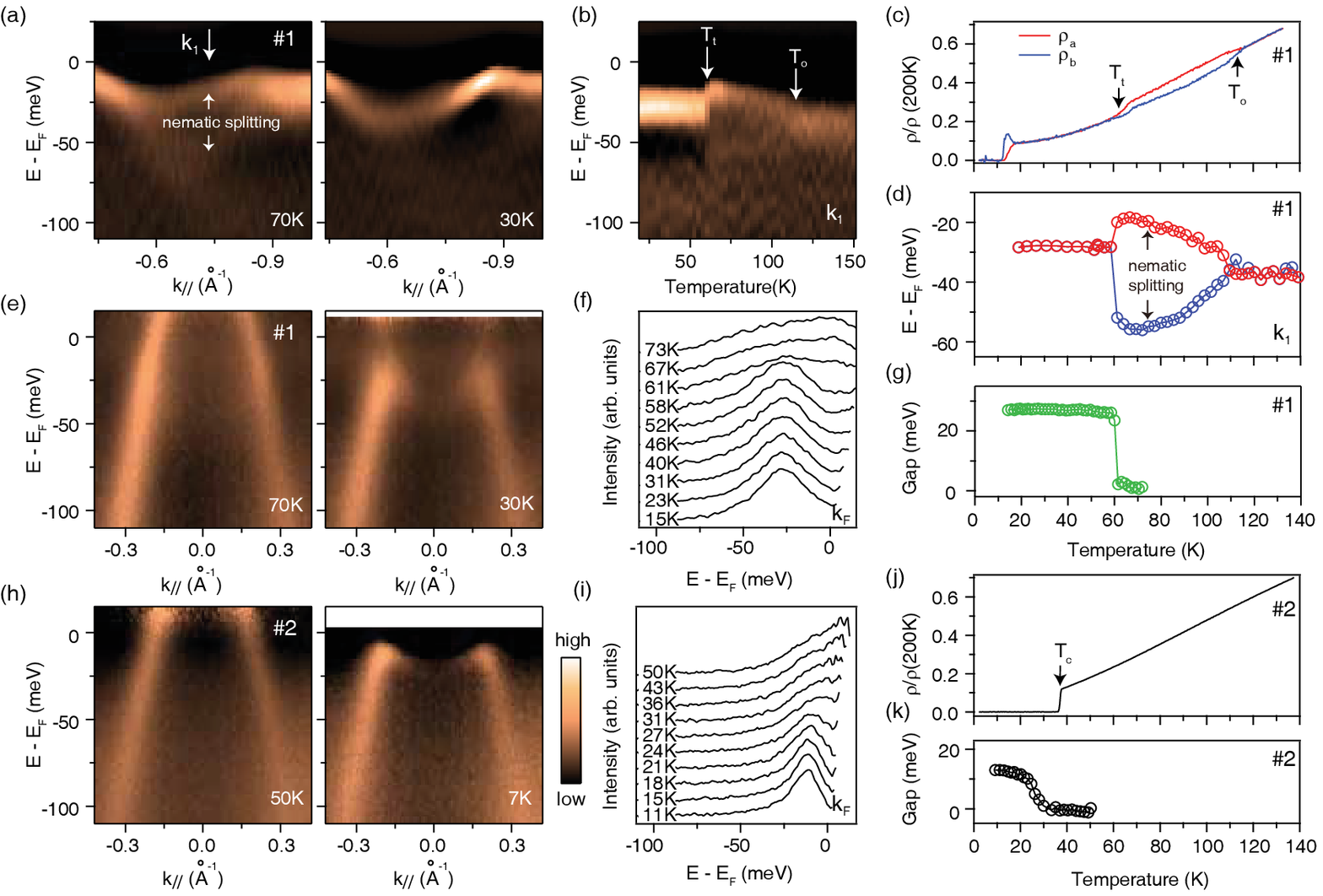}
\caption{ (color online) Characterization of the phase transitions using ARPES and resistivity measurements. (a) Second derivative images of the energy-momentum cuts taken along the $\Gamma$-M direction at 70~K and 30~K in the sample $\#$1. (b)Second derivative merged image of the energy distribution curves (EDCs) taken at $k_1$ in (a). The AFM/O transition temperatures ($T_o$) and AFM/T transition temperature ($T_t$) are illustrated using white arrows. (c) Temperature dependence of the in-plane resistivity anisotropy taken on a detwinned sample. (d) Temperature dependence of the peak positions taken from the data in (b). (e) Energy-momentum cuts taken around the $\Gamma$ point at 70~K and 30~K in the sample $\#$1. (f) The temperature dependence of the EDCs taken at the Fermi crossing ($k_F$) in (e). The EDCs were divided by a Fermi-Dirac function to remove the Fermi energy cut-off. (g) The temperature dependence of the energy gap determined using the peak positions in (f). (h and i) are the same as (e and f), but taken in the sample $\#$2. (j) Temperature dependence of the normalized resistivity taken in the sample $\#$2. (k) The temperature dependence of the superconducting gap determined using the peak positions in (i). All ARPES data were taken using 25~eV photons. }\label{f2}
\end{figure*}

We synthesized Sr$_{1-x}$Na$_x$Fe$_2$As$_2$ single crystals with two doping levels $x$ = 0.36 ($\#$1) and $x$ = 0.49 ($\#2$) [Fig.~\ref{f1}(a)] \cite{gap20} using self-flux method. Mixtures of Sr, NaAs and FeAs in a molar ratio of 0.64~:~$y$~:~4 were loaded in alumina crucibles, and sealed in iron crucibles under the Ar atmosphere. The iron crucibles were heated to 1100~$^\circ$C and kept for 12~h, and then slowly cooled down to 800~$^\circ$C at the rate of 1.5~$^\circ$C/h before the furnace was shut down. The molar ratio of $y$ varied from 2.5 to 4 for different chemical composition. The actual chemical composition of $x$ was measured using scanning electron microscope (SEM) equipped with an energy dispersive x-ray (EDX) analyzer. The resistivity measurements were carried out on a Quantum Design physical property measurement system (PPMS) using a standard four-probe method. The AFM/T transition was observed in all measured samples with a small deviation of transition temperature less than 10~K \cite{supp}. The resistivity anisotropy measurements were carried out on a PPMS system using Montgomery method and a self-developed detwinning device \cite{supp}. ARPES measurements were performed at Stanford Synchrotron Radiation Light source (SSRL) Beamline 5-4 and Peking University. The photon energy is 21.2~eV for the experiments at Peking University and 25~eV for the experiments at SSRL. The overall energy resolution was $\sim$8~meV and the angular resolution was $\sim$0.3$^{\circ}$. All the samples were measured in ultrahigh vacuum with a base pressure better than 6$\times$10$^{-9}$~Pa.

The Fermi surface of iron-based superconductors consists of three hole pockets at the Brillouin zone center ($\Gamma$) and two electron pockets at the Brillouin zone corner (M) \cite{gap1, gap2, gap24}. The low-energy electronic structure is constructed by the $d_{xz}$, $d_{yz}$ and d$_{xy}$ orbitals [Fig.~\ref{f1}(b)]. In the AFM/T phase, due to translational symmetry breaking, the Brillouin zone reduces in size and bands fold between $\Gamma$ and M. SDW gaps open at the band crossings between the hole and electron bands [Fig.~\ref{f1}(c)]. We first characterized the phase transitions using APRES and resistivity measurements. According to previous ARPES studies, the electronic signature of rotational symmetry breaking is the nematic band splitting near the M point \cite{gap17, gap25,gap26}. In the sample $\#$1, the nematic band splitting first emerges at around 110~K and then suddenly disappears at 60~K [Fig.~\ref{f2}(a), \ref{f2}(b), and \ref{f2}(d)]. This is fully consistent with the resistivity measurement, where the resistivity anisotropy emerges in the AFM/O phase and vanishes in the AFM/T phase [Fig.~\ref{f2}(c)]. The disappearance of both the nematic band splitting and resistivity anisotropy confirms that the four-fold rotational symmetry recovers in the AFM/T phase. In the spectra taken at the $\Gamma$ point [Fig.~\ref{f2}(e)], one hole-like band crosses the Fermi energy ($E_F$) at 70~K. When entering the AFM/T phase, an electron-like band emerges indicating a band folding due to transitional symmetry breaking. A SDW gap opens at the Fermi crossings with a gap size around 27~meV [Fig.~\ref{f2}(f) and \ref{f2}(g)]. The sharp opening of the SDW gap is consistent with the first-order character of the AFM/T phase transition. Note that, the resistivity measurement shows that the sample $\#$1 further enters the superconducting phase at around 12~K. The superconducting gap is too small to be resolved at our lowest achievable experimental temperature (7~K). The tiny $T_c$ difference between $\rho_a$ and $\rho_b$ could be due to the drifting of uniaxial pressure during the resistivity measurement \cite{supp}. On the other hand, in the sample $\#$2, the resistivity shows a linear temperature dependence, which could be explained by the non-Fermi liquid behavior observed in optimal-doped iron-based superconductors \cite{nonfermi}. The superconducting transition occurs at around 35~K [Fig.~\ref{f2}(j)]. The superconducting gap opens gradually at $T_c$ with a gap size around 10~meV [Fig.~\ref{f2}(i) and \ref{f2}(k)]. The superconducting coherence peak emerges in the superconducting phase whose width is limited by the experimental energy resolution.

\begin{figure*}[t]
\includegraphics[width=15cm]{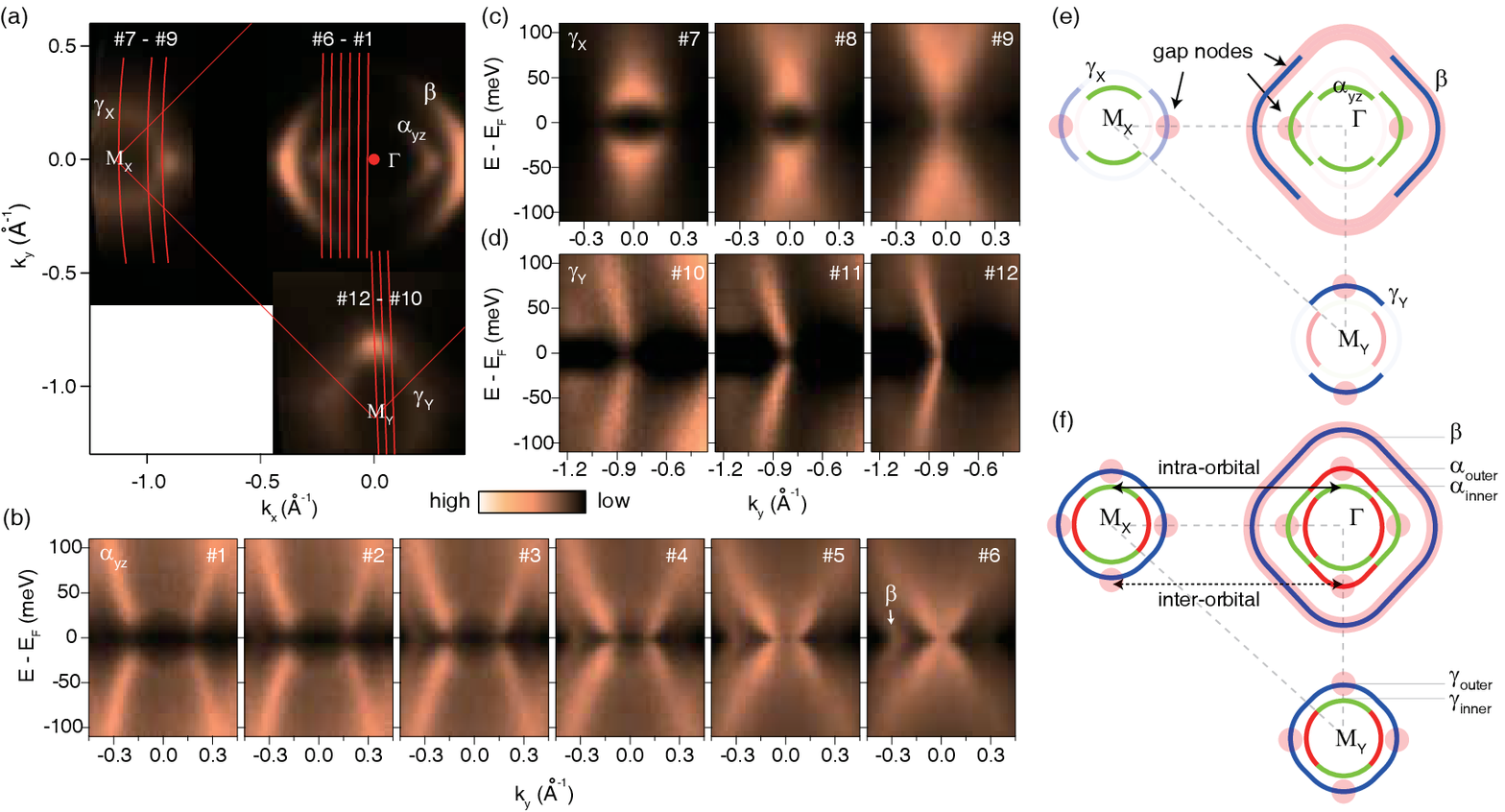}
\caption{(color online) The SDW gap structure in the AFM/T phase. (a) Fermi surface mappings taken in the AFM/T phase. (b) The symmetrized images of the energy-momentum cuts taken around the $\Gamma$ point. The cut momenta are illustrated in (a). The data were taken using 25~eV photons with linear vertical polarization. (c and d) are the same as (b) but taken on the electron pockets around the M$_X$ and M$_Y$ points, respectively. The data were taken using 21.2~eV photons with mixed polarization. (e) The summarized gap anisotropy on all observable sections of Fermi surface under our experimental setup. (f) The recovered gap anisotropy of all Femi surface sheets. The solid line shows the intra-orbital scattering between the $\alpha_{inner}$ and $\gamma_{inner}$ pockets. The dotted line shows the inter-orbital scattering between the $\alpha_{outer}$ and $\gamma_{outer}$ pockets. All data were taken on sample $\#$2 at 15~K.}\label{f3}
\end{figure*}

We then measured the SDW gap distributions on all Fermi surface sheets in the AFM/T phase [Fig.~\ref{f3}]. We note that the SDW gap is not strictly particle hole symmetric. Because ARPES can only detect the occupied states, we determined the energy gap size using the energy difference between the dispersion minima and the $E _F$. The image symmetrization is then used to illustrate the gap anisotropy in the AFM/T phase. First, the SDW gap is strongly anisotropic on the inner hole pocket ($\alpha_{yz}$) [Fig.~\ref{f3}(b)]. From cut $\#$1 to $\#$6, the gap magnitude decreases when moving away from the vertical high symmetry direction. Gap node forms on the horizontal corner of the $\alpha_{yz}$ hole pocket (Cut $\#$6). There is no energy gap opening on the outer hole pocket ($\beta$) [Fig.~\ref{f3}(b)]. For the electron pockets, the SDW gap is anisotropic on the $\gamma_X$ electron pockets at the M$_X$ point [Fig.~\ref{f3}(c)]. The gap magnitude decreases when going from cut $\#7$ to $\#9$. Gap nodes locate on the horizontal corner of the $\gamma_X$ electron pocket. At the M$_Y$ point, the gap anisotropy rotates 90 degrees with the gap nodes along the vertical high symmetry direction [Fig.~\ref{f3}(d)].

The multi-band electronic structure of iron-based superconductors challenges the experimental measurement of gap structure. For example, the existence of double peak features has been reported in Ba$_{1-x}$K$_x$Fe$_2$As$_2$, which were attributed to the nearly degenerated $d_{xz}$ and $d_{yz}$ bands \cite{BaK1}. Here, unitizing linear vertical polarization, we could reduce the number of observable bands by half as shown in Fig.~\ref{f3}(e) and \ref{f3}(f) \cite{gap27, gap28, supp}. Around the $\Gamma$ point, we only probe the $d_{yz}$ sections of the inner hole pockets while around the M point, due to the glide-mirror symmetry of the iron-arsenic plane, only the $d_{xz}$/$d_{xy}$ and $d_{yz}$/$d_{xy}$ electron pockets show up at the M$_X$ and M$_Y$ points respectively. As a result, all energy-momentum cuts shown in Fig.~\ref{f3} consists of only one prominent band, while all other bands including the folded bands are either weak or absence.

We then discuss the SDW gap structure of the AFM/T phase. It is worthy to note that, while the quantitative gap magnitude may be affected by the matrix element effect and weak shadow bands, the gap angular symmetry, nodal locations and relative gap magnitude are all robust. The gap anisotropies show a common C2 rotational symmetry on the $\alpha_{yz}$, $\gamma_X$ and $\gamma_Y$ pockets, suggesting an inter-pocket nesting among them. Under the Fermi surface nesting scenario, the nesting is most effective when the nesting portions of Fermi surface have the same orbital character \cite{gap29, gap30}. By mapping the measured gap structure to the original Fermi surface [Fig.~\ref{f3}(f)], the correlation between the nesting condition and gap magnitude is obvious. First, the pocket size of the $\beta$ hole pocket is large and there is no corresponding nested electron pocket. Therefore, no energy gap opens on the $\beta$ hole pocket. Second, the $\alpha_{inner}$ hole and $\gamma_{inner}$ electron pockets share similar orbital characters and pocket sizes. The intra-orbital scattering between them results in large and isotropic energy gaps on both the $\alpha_{inner}$ and $\gamma_{inner}$ pockets. Third, for the $\alpha_{outer}$ hole and $\gamma_{outer}$ electron pockets, their different orbital characters result in an inter-orbital nesting between them. Gap nodes exist along the high-symmetry directions where the orbital characters of bands are well defined. Note that, the applicability of the Fermi surface nesting scenario does not necessarily mean that the magnetism originates purely from the Fermi surface nesting of itinerant electrons. Previous experiments show that the magnetism in iron-based superconductors shows both the local and itinerant characteristics \cite{gap29}. Besides the nesting condition, the spin-orbital coupling (SOC) can also lead to a gap anisotropy \cite{gapSOC}. However, the strong orbital-selectivity and gap nodes observed here are inconsistent with a SOC-related gap anisotropy.

\begin{figure}[t]
\includegraphics[width=8.7cm]{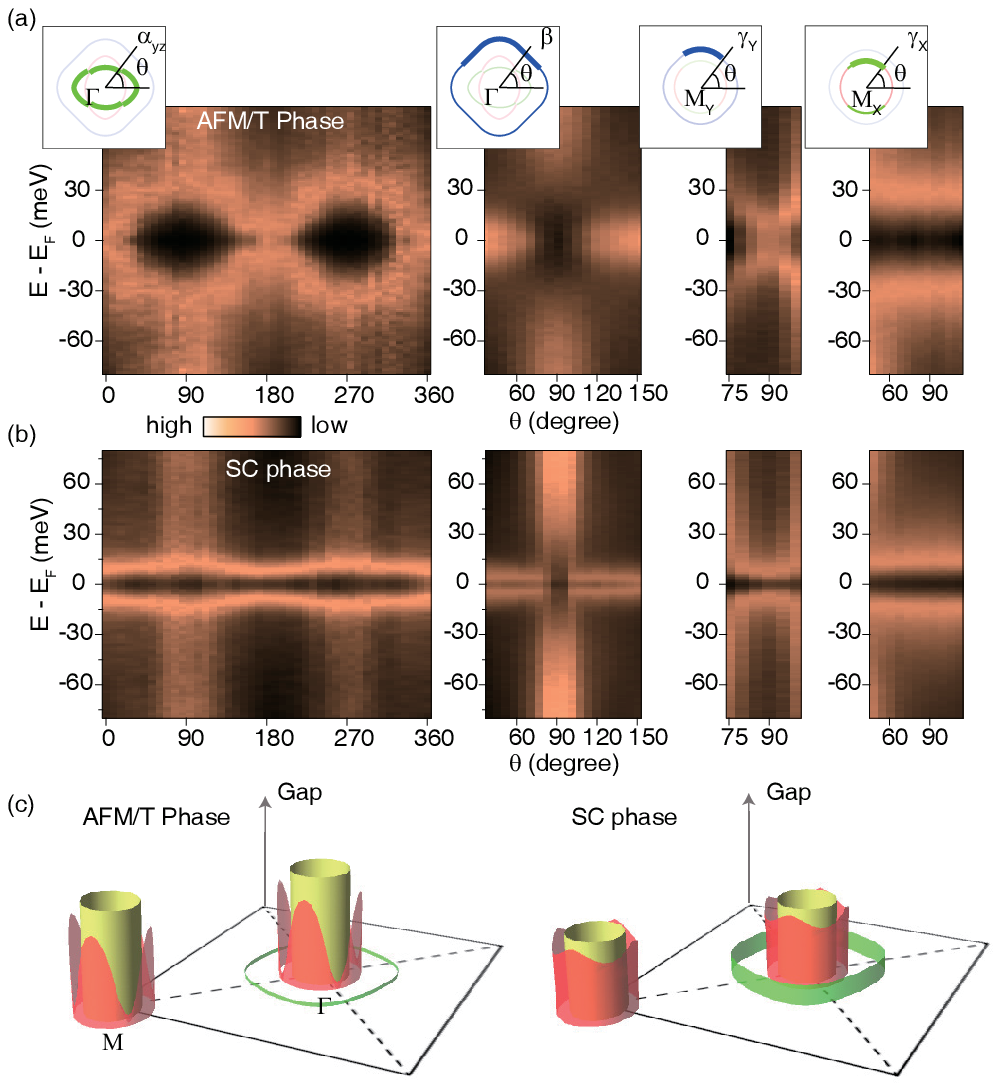}
\caption{ (color online) The comparison of gap anisotropies taken in the AFM/T and superconducting phases. (a) The SDW gap distributions on different Fermi surface sheets taken in the AFM/T phase on sample $\#$1 at 15~K. The EDCs are selected at the gap minima, which are determined using the Fermi surface mapping, EDCs and MDCs \cite{supp}. The EDCs are merged into images to better visualize the gap anisotropy. The measured momenta are shown in the inset panels. (b) The schematic of the SDW gap anisotropies in the AFM/T phase. (c and d) are the same as (a and b), but taken in the superconducting phase on sample $\#$2 at 15~K. }\label{f4}
\end{figure}
With the same experimental setup, we measured the superconducting gap anisotropy in the superconducting sample $\#$2, and compared it with the SDW gap anisotropy in Fig.~\ref{f4}. We found that the superconducting gap anisotropy is isostructural to the SDW gap anisotropy. For the $\alpha_{yz}$ hole pocket, the superconducting gap anisotropy is two-fold symmetric with the gap maxima along 90 degrees direction and gap minima along 0 degrees direction. The superconducting gap magnitude is relatively small on the $\beta$ hole pocket. For the electron pockets, the superconducting gap minima locate along the 90 degrees direction on the $\gamma_Y$ electron pocket, while the energy gap is large and isotropic on the $\gamma_X$ electron pocket.

The energy gap anisotropy in the momentum space reflects the detailed form of interactions. For example, in cuprates, both the AFM and superconducting gaps show maxima at the anti-nodal region, reflecting the importance of the nearest-neighbour exchange interaction in both the AFM and superconducting phases. Similarly here in iron-based superconductors, the SDW and superconducting gap maxima all locate at the inner pockets close to the $\Gamma$ and M points, suggesting the importance of ($\pi$, 0) scattering and next-nearest neighbour exchange interaction in both the SDW and superconducting phases. Furthermore, the origin of the pairing interaction in iron-based superconductors is still under debate. Here, our observation of the isostructural gap anisotropies with similar angular and pocket dependencies establishes a close connection between the magnetism and superconductivity in iron-based superconductors, strongly suggesting that the superconducting pairing is mediated by spin fluctuation.

Our results show that the SDW and superconducting gap magnitude correlates closely with the orbital character of Fermi surface. While the gap minima are observed on the $d_{xy}$ pockets, the gap maxima locate on the $d_{xz}$/$d_{yz}$ pockets in both the AFM/T and superconducting phases. This result show the dominating role of intra-orbital scattering, and point out the existence of strong orbital-selectivity in both the superconducting and magnetic phases in iron-based superconductors. In FeSe superconductor, both scanning tunneling microscope (STM) and ARPES found a close correlation between the superconducting gap magnitude and orbital character of Fermi surface suggesting an orbital-selective superconducting pairing \cite{gap31, gap32}. For the magnetism, it has been proposed that the magnetism may originate from either the $d_{xz}$/$d_{yz}$ or $d_{xy}$ orbitals depending on the detailed electronic structure near the Fermi energy \cite{gap33}. Further theoretical and experimental studies are required to understand how different orbitals play roles in constructing the pairing and magnetic interactions in iron-based superconductors. For example, an orbital-selective electronic correlation has been proposed in iron-based superconductors, pointing out the coexistence of both local and itinerant orbitals \cite{gap33}. An orbital-dependent pairing symmetry has been proposed where the phases of the $d_{xz}$/$d_{yz}$ and $d_{xy}$ pairing channels are opposite.

We gratefully thank Yuan Li for helpful discussions. This work is supported by National Natural Science Foundation of China (Grant No.~11888101, No.~91421107, and No.~11574004), the National Key Research and Development Program of China (Grant No.~2016YFA0301003 and No.~2018YFA0305602). Stanford Synchrotron Radiation Lightsource is operated by the Office of Basic Energy Sciences, U.S. Department of Energy.

\end{document}